\documentclass{nature}
\usepackage{graphicx}
\makeatletter
\let\saved@includegraphics\includegraphics
\AtBeginDocument{\let\includegraphics\saved@includegraphics}
\renewenvironment*{figure}{\@float{figure}}{\end@float}
\makeatother

\usepackage{multirow}
\usepackage{subfigure}
\usepackage{mathtools}
\usepackage[colorlinks=true,linkcolor=blue,citecolor=black,urlcolor=blue]{hyperref}
\usepackage{soul,xcolor}
\usepackage{authblk}
\usepackage{lineno}
\usepackage[labelfont=bf]{caption}
\usepackage[figurename=Figure ]{caption}

\begin{document}

\title{Universality of defect-skyrmion interaction profiles}
\author{Imara Lima Fernandes, Juba Bouaziz, Stefan Bl\"{u}gel \& Samir Lounis}
\affil{Peter Gr\"{u}nberg Institut and Institute for Advanced Simulation, Forschungszentrum J\"{u}lich and JARA, D-52425 J\"{u}lich, Germany}

\maketitle

\begin{abstract}
Magnetic skyrmions are prime candidates for future spintronic devices. However, incorporating them as an entity for information technology hinges on their interaction with defects ubiquitous to any device. Here we map from first-principles, the energy-profile of single skyrmions interacting with single-atom impurities, establishing a generic shape as function of the defect's electron filling. Depending on their chemical nature, foreign 3d and 4d transition metal adatoms or surface-implanted defects can either repel or pin skyrmions in PdFe/Ir(111) thin films, which we relate to the degree of filling of bonding and anti-bonding electronic states inherent to the proximity of the non-collinear magnetic structure. Si\-mi\-la\-rities with key concepts of bond theories in catalysis and surface sciences imbue the universality of the shape of the interaction profile and the potential of predicting its interaction. The resulting fundamental understanding may give guidance for the design of devices with surface-implanted defects to generate and control skyrmions.
\end{abstract}

\maketitle

Magnetic skyrmions\cite{Bogdanov1989,Roessler2006} are non-collinear spin textures with particle-like properties, which are currently under intense scrutiny due to the rich science and technological potential for future information and communication devices\cite{Fert2013,Fert2017,Sampaio2013,Tomasello2013,Kang2016,Zhang2015b,Crum2015,Zhou2014,Zhang2015,Kim2014,Finocchio2015}. Skyrmions are stabilized in a plethora of materials\cite{Fert2017,Wiesendanger2016} and their topological nature\cite{Roessler2006,Skyrme1962,Nagaosa2013} can provide an energy-barrier stabilizing them up to room temperature\cite{Jiang2015,Woo2016,Boulle2016,Moreau-Luchaire2016}. The low current densities estimated to move them in a racetrack memory\cite{Fert2013} have raised expectations for the development of new technologies with low energy consumption. Requirements of future power efficient information technology such as high density, high speed information processing and storage translate into additional challenges for skyrmion research. Consequently, one focus is on sub-10~nm skyrmions, a second one is on high retention, high reliability of devices and fast skyrmion motion. The latter hinge on material inhomogeneities and defects. For example, the skyrmion motion is realized with rather low currents in films of B20 compounds \cite{Jonietz2010,Yu2012}, detailed experiments of interface stabilized skyrmions in ultrathin heavy metal/ferromagnetic bilayers and multilayers demonstrated a complex dynamical behaviour as function of applied currents\cite{Woo2016,Legrand2017,Jiang2016,Litzius2017}, with low skyrmion velocities. In contrast to phenomenologically-based predictions\cite{Fert2013,Sampaio2013,Iwasaki2013a,Iwasaki2013b}, their nucleation, retention, motion and velocity are strongly affected by material inhomogeneities\cite{Woo2016,Jiang2015,Jiang2016,Litzius2017,Hanneken2016}. The latter are paramount in influencing the depinning currents and subsequently the skyrmion-velocity versus applied current efficiency. Furthermore, the skyrmion Hall angle\cite{Nagaosa2013} is surprisingly contingent on the skyrmion velocities, probably, due to the competition between the pinning potentials of the defects and the skyrmion driving force\cite{Jiang2016}.

Interestingly, interfacial defects have been noticed at the vicinity of skyrmions written in PdFe bilayers on the Ir(111) surface after injection of an electric current from a scanning tunneling microscope (STM)\cite{Romming2013}. This indicates that their presence is needed to pin the nucleated skyrmions in order to be observed. While it is not obvious how to ease out material inhomogeneities naturally present in devices, control and design of defective surfaces is an appealing strategy to manipulate skyrmions. First successful attempts were realized recently. For instance, constrictions are built to generate skyrmion bubbles\cite{Jiang2015} and  skyrmions nucleated in  PdFe/Ir(111)  could be moved with a Co trimer deposited on the surface but surprisingly not with a single Co adatom\cite{Hanneken2016}. Several theoretical studies based on phenomenological assumptions have been conducted to prospect the impact of defects (see e.g.\cite{Fert2013,Sampaio2013,Liu2013,Mueller2015,Stosic2017}) and a recent ab-initio study addressed the case of a lattice of skyrmions with a periodic arrangement of a few defects in bulk MnSi\cite{Choi2016}, yet the energy-profile of  interactions between single magnetic skyrmions and defects, essential to all the mentioned effects, requesting a realistic description of their electronic structure and its fundamental understanding is terra incognita.

Here we address from first-principles (see Methods section) the impact of single atomic defects  on single magnetic skyrmions generated in a fcc-PdFe bilayer deposited on the Ir(111) surfaces. These defects include the full series of transition metal atoms: 3d (Sc, Ti, V, Cr, Mn, Fe, Co, Ni), 4d (Y, Zr, Nb, Mo, Tc, Ru, Rh) as well as Cu and Ag atoms. They are located either on top of (adatoms) or embedded in (inatoms) the Pd surface layer (see Figure~\ref{Fig1}a). Adatoms can be manipulated with STM, while inatoms are generated naturally from interfacial intermixing processes or by ion-implantation\cite{Persaud2005}. The chosen substrate is well characterized\cite{Crum2015,Romming2013,Dupe2014,Simon2014} and hosts isolated N\'eel-type skyrmions with a diameter of 1-5 nm pinned probably at interface defects and surrounded by a ferromagnetic background in an external magnetic field.

\section*{Results}
\subsection{Impact of defects on the non-collinearity.}

\begin{figure}
	\includegraphics[width=\columnwidth]{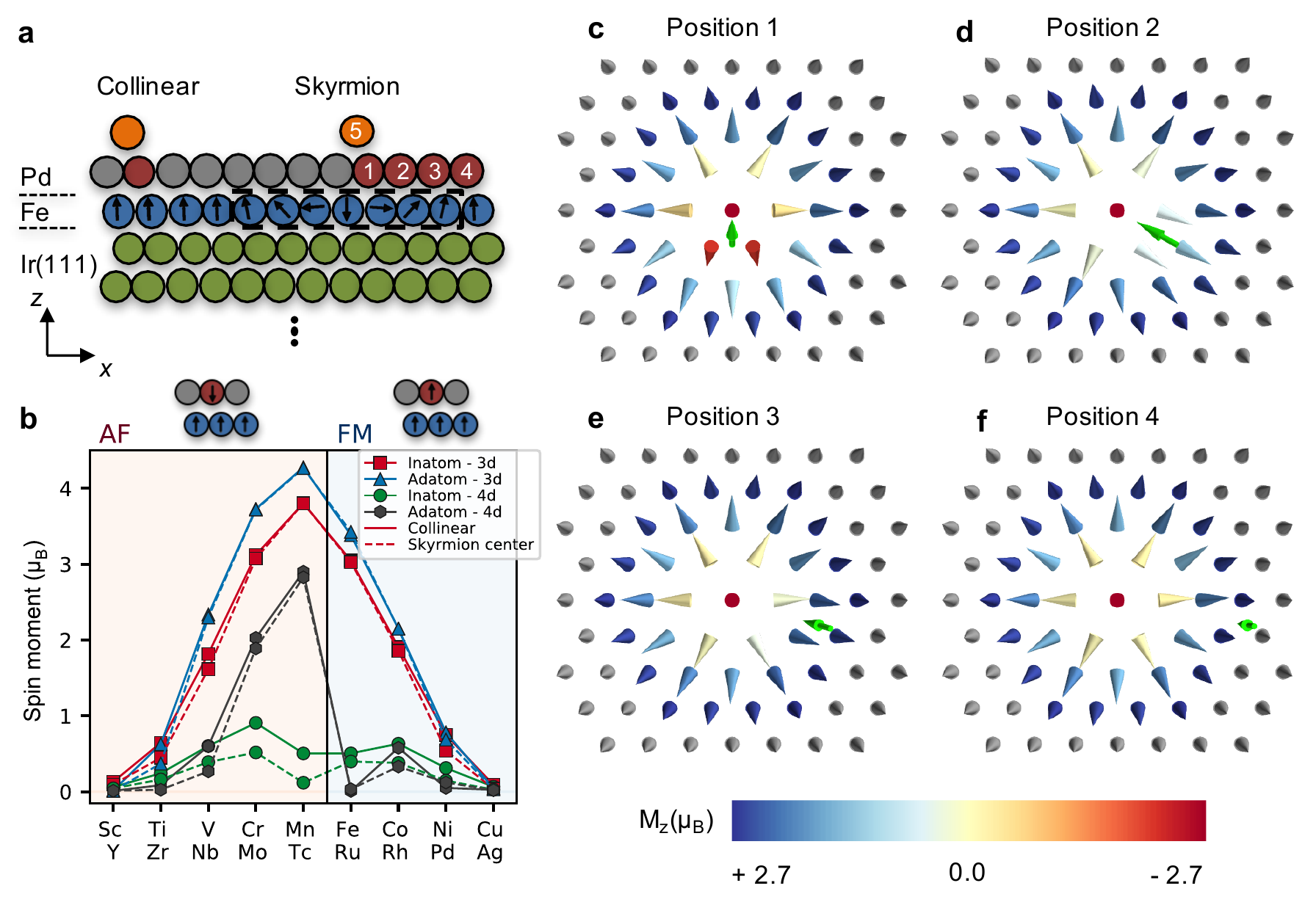} 
	\caption{\textbf{Skyrmions and single atomic defects: spin structure and magnetic moments.} (\textbf{a}) Illustration of the cross-section of PdFe/Ir(111) surface with the defects location (positions label 1 to 5). \textbf{(b)} Chemical trend of local magnetic moments as function of the tran\-si\-tion-metal impurity as single defects located at the collinear configuration (solid line) or close to the skyrmion center (dashed line). The magnetic coupling between the defect and the Fe-substrate is antiferromagnetic (AF) on left of Fe and ferromagnetic (FM) on right of Mn. (\textbf{c-f}) Spin structure of a skyrmion interacting with V-inatom at different positions. The green arrow represents the direction of the impurity magnetic moment and the colour bar represents the magnitude of the z-component of the magnetization for each Fe atom.}
	\label{Fig1}
\end{figure}

The nature of the magnetic coupling between the defects and the Fe-layer is settled by the electronic hybridization. As commonly known, we find the magnetic moments of early transition elements couple antiferromagnetically to the ones of the PdFe bilayer whereas the late ones couple ferromagnetically (see e.g.\cite{Dederichs1998,Dederichs1998_PRB,Lounis2005}). Also, the magnetic moments for 3d inatoms, 3d and 4d adatoms follow Hund's first rule (Figure \ref{Fig1}b) and are largest at half-filling. In contrast, the magnetic moments of 4d inatoms follow a more band-type behaviour with a subtle dependence on the band-filling and are strongly sensitive to the magnetic behaviour of the neighbouring atoms. The latter can reshape the electronic structure of 4d inatoms and, consequently, their magnetic properties. 
 When located in the Pd layer, the 4d magnetic moments are mainly induced by the underlying Fe atoms because of direct hybridization of their electronic states. Since 4d orbitals extend further in space than 3d orbitals, they are more responsive to their electronic environment. Close to the skyrmion core, due to the non-collinearity, the effective induced moment is smaller than the one from the ferromagnetic background (see Figure \ref{Fig1}b). When deposited atop the substrate, the coordination is reduced, modifying the electronic structure such that several of 4d elements follow the atomic picture described by Hund's first rule.

 As an example, we show in Figures \ref{Fig1}c-f a skyrmion with a diameter of 2.2 nm interacting with a V-inatom at different locations. As found for other impurities, the skyrmion profile is non-trivially reshaped and the locally induced spin-stiffness depends on the impurity-substrate magnetic exchange interaction and the magnitude of the impurity moments. Despite the asymmetric shape, the skyrmion topological charge\cite{Nagaosa2013} remains protected for all elements and positions. The skyrmion recovers its usual high symmetry once its center is moved away from the impurity, Figures \ref{Fig1}e-f.

\subsection{Defect-skyrmion interaction energy-profiles.}

\begin{figure}
	\includegraphics[width=\columnwidth]{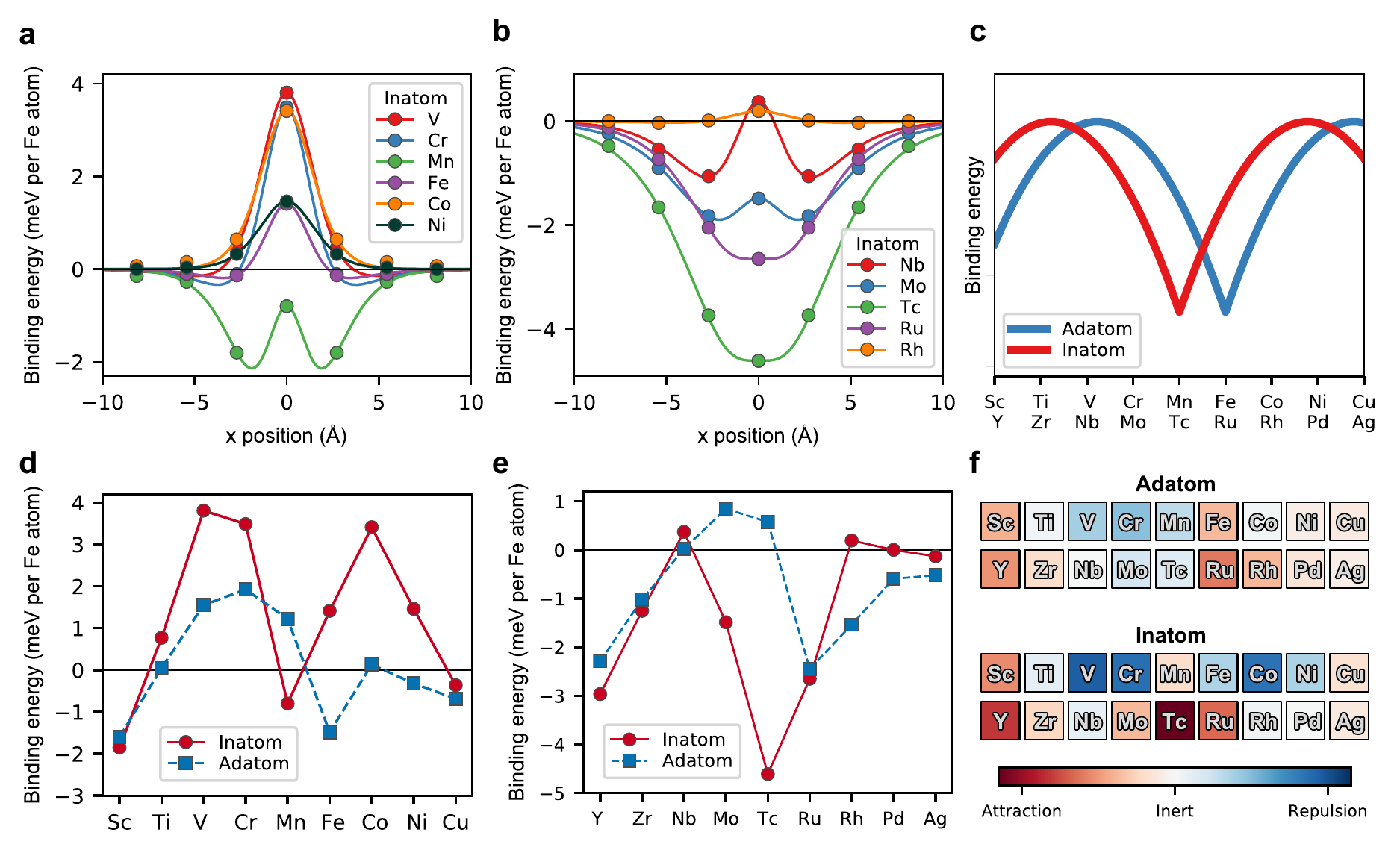} 
	\caption{\textbf{Skyrmion-defect binding energies}. (\textbf{a-b}) Binding energies as function of the $\text{x}$ position of the inatoms with respect to the skyrmion core as defined in Figure \ref{Fig1}. A negative (positive) binding-energy indicates an attractive (repulsive) skyrmion-defect interaction. The discrete \textit{ab initio} results (circles) are interpolated with lines resulting from fits to Morse potentials. (\textbf{c}) Schematic illustration of the M-shape behaviour for the interaction profile with a two-peak feature close to the edges of the transition metal series and a minimum around half-filling, i.e. in the middle of the series. (\textbf{d-e}) Impact of band-filling on the  binding energies considering inatom (solid) and adatom (dashed)  impurities located close to the skyrmion center.  (\textbf{f}) Periodic table for inatom and adatom defects with the colour scale indicating the strength of their binding energies. Red corresponds to attraction and blue to repulsion.}
	\label{Fig2}
\end{figure}

The binding energy between the skyrmion and the inatom impurities, plotted in Figure~\ref{Fig2}a-b, is obtained from energy differences considering two setups where the skyrmion and defect are either close to each other or far apart. A negative binding-energy expresses an attractive impurity-skyrmion interaction; positive binding-energy, repulsive interaction. One notices that the chemical nature of the defects impacts non-trivially on the impurity-skyrmion binding-energy leading to interaction profiles with barrier- or well-like shapes with different widths and heights or depths, which will then affect the motion of skyrmion at the vicinity of a defect.

For conciseness, we mainly focus our analysis on the atoms embedded in the Pd-layer. Inatoms of the 3d series tend to repel the skyrmion in contrast to those of the 4d series, which act as pinning centers. Exceptions are Sc, Mn and Cu-inatom, which can attract a skyrmion, and a rather inert Rh-inatom. That means a skyrmion can move around a Rh-inatom without being captured or repelled. While the energy-barriers are centered close to the core of the skyrmion, the energy-wells can be slightly off-centered as found for Mn and early elements of the 4d series (Nb, Mo). Furthermore, the widths of the energy-profiles are largest for the 4d impurities, probably due to the large spatial extension of their 4d orbitals. Thus, hybridization with the substrate electronic states is stronger for the 4d impurities than for the 3d ones.

After mapping the magnetic exchange interactions from \textit{ab initio} to a generalized Heisenberg model, important insights can be deduced. Even though the model can suffer from restricting the magnitude of the magnetic moments to change upon rotation, we obtain qualitatively similar results (for more details see Supplementary Note 1). We learn that impurities lower the magnetic exchange interactions within the Fe-layer facilitating the creation of a magnetic skyrmion. Moreover, the impurities can be of two types: (i) strongly magnetic, providing a strong magnetic exchange interaction stiffening locally the skyrmion, which leads to repulsion and (ii) weakly magnetic, behaving like Pd, where the substrate contribution prevails and leads to pinning. The magnetic interactions are closely related to the details of the electronic structure, discussed in the next paragraphs.

When plotted as function of the atomic number (Figure~\ref{Fig2}d-e), the binding energy of the 3d and 4d defects exhibits an M-shape, plotted schematically in Figure~\ref{Fig2}c, with a two-peak feature close to the edges of the transition metal series and a minimum around half-filling, i.e. in the middle of the series. For instance, when moving from left to right across the 3d (4d) row in the periodic table, the initially negative binding energy reaches the first maximum at V(Nb)-inatom followed by a minimum at half-filling, i.e. Mn(Tc)-inatom. The second maximum is observed at Co(Rh)-inatom, after which the binding energy decreases. Interestingly, when moving the impurities from the inatom to the adatom location, the energy profiles seem to shift one column to the right in the periodic table. In the 3d-series, the minimum is found close to half-filling, i.e. at Fe-adatom, while it is located at Ru-adatom instead of Tc-adatom in the 4d-series. Thus with such a shift, the second maximum, previously close to the full-filling of the d-states, moves to the edge of the transition metal series as shown schematically in Figure~\ref{Fig2}c. This common behaviour across the 3d and 4d series is a typical signature of a band-filling effect and is qualitatively similar to the trend of cohesion\cite{Friedel1977,Blandin1965} or surface energies\cite{Dederichs1998,Dederichs1998_PRB} of transition metals. It suggests that shapes of the defect-skyrmion interaction profile can be predicted just by knowing the emplacement of their constituent elements in the periodic table. A portion of the latter is established in Figure~\ref{Fig2}f with colours indicating the strength and nature of the skyrmion-defect binding energy from which design principles for skyrmion functionality can be derived.

\subsection{Electronic structure of inatoms at the vicinity of skyrmions.}

In contrast to 4d-elements, 3d-elements have the strongest exchange splitting, which essentially leads to a discrepancy of the skyrmion-defect interactions. We illustrate the electronic mechanisms behind this trend by analyzing the d-projected local density of states (LDOS) of two impurities: the strongest pinning center, Tc-inatom, and the strongest repelling defect, V-inatom, whose magnetic moments couple antiferromagnetically to the substrate. We limit our discussion to two impurity locations: close to the skyrmion center and on the ferromagnetic background away from the skyrmion.

\begin{figure}
	\includegraphics[width=\columnwidth]{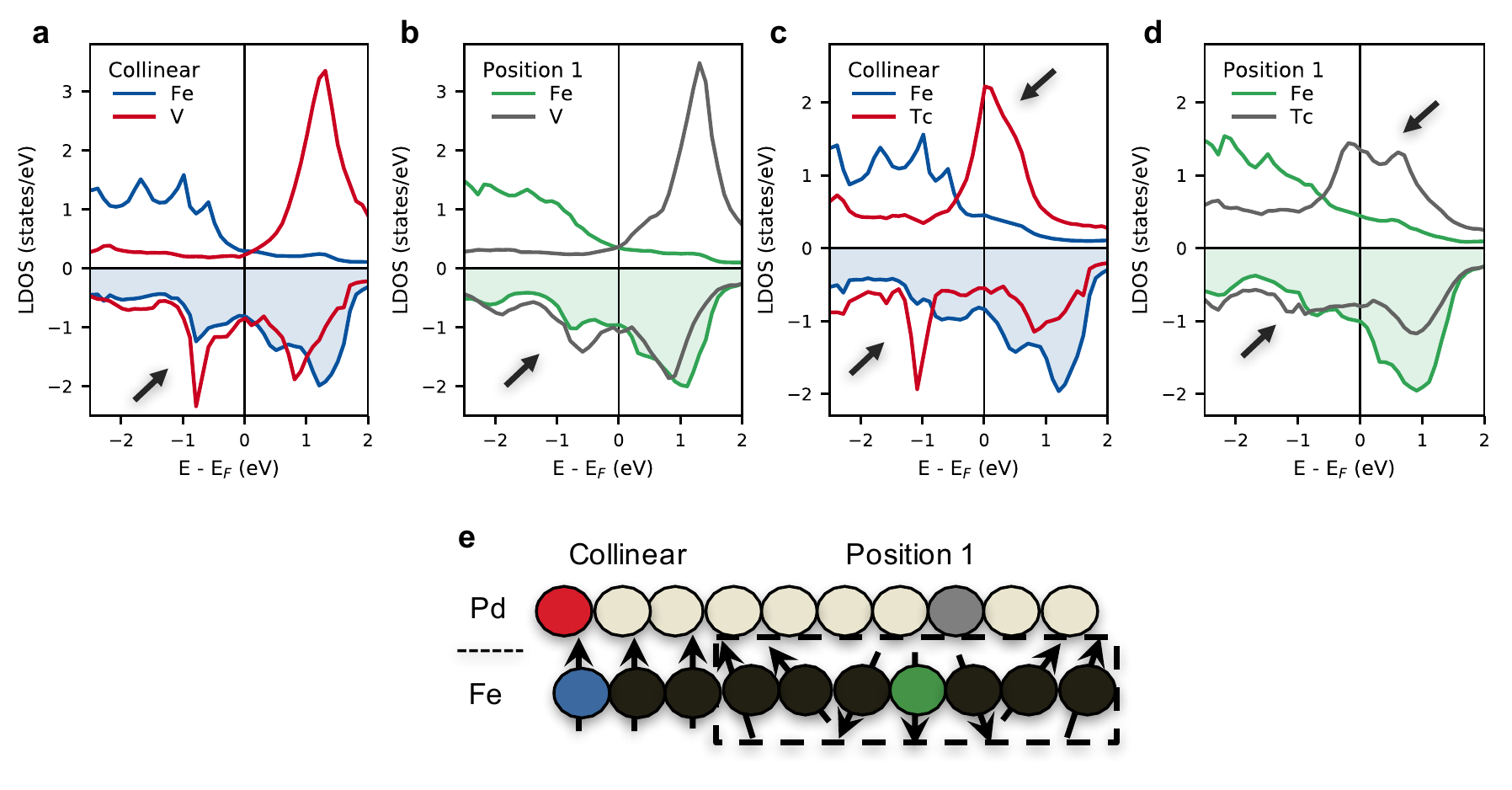} 
	\caption{\textbf{Impact of magnetic environment on the impurities electronic structure.} (\textbf{a-b}) The local density of states (LDOS) of V-inatom, the most skyrmion repelling impurity, as obtained when located within the collinear region (red lines in \textbf{a}) or close to the skyrmion core (grey lines in \textbf{b}). (\textbf{c-d}) Analogously for the most attracting impurity, Tc-inatom. For comparison, the LDOS of one of the closest Fe atoms to the impurity is plotted. Green corresponds to the skyrmion-case whereas blue to the ferromagnetic background. (\textbf{e}) Illustration of atom position corresponding to colour code used in \textbf{a-d}.}
		\label{Fig3}
\end{figure}

The sharpness of the minority-spin V-resonance can be explained in terms of hybridization and its large exchange splitting. In the collinear configuration, the minority-spin virtual bound state (VBS) is unoccupied (Figure~\ref{Fig3}a, red curve) and weakly  hybridizes with the almost fully occupied majority-spin band of Fe (Figure~\ref{Fig3}a, blue curve). However, the majority-spin VBS of V-inatom hybridizes with the Fe minority-spin states generating a large band (see arrow in Figure~\ref{Fig3}b) with clear bonding and anti-bonding states. The interplay between the number of valence electrons and exchange splitting coupled to hybridization processes give rise to the specific dependence on the band filling. The rather prominent bonding state is occupied which provides stability to the collinear configuration. When moving the skrymion core towards the V-inatom, the non-collinearity leads the opening of additional hybridization channels due to spin-mixing of the Fe states. That broadens the occupied bonding-state and drastically lowers its magnitude (see arrow in Figure~\ref{Fig3}b, grey curve), which makes this configuration unfavourable. In other words, V-inatom repels the skyrmion.

The majority-spin band of V-inatom and Tc-inatom (lower pannel in Figures~\ref{Fig3}a-c, red curve) are  rather similar. However, the spin-minority VSB of Tc shifts to the Fermi energy due to the weaker exchange splitting (see upper arrow in Figure~\ref{Fig3}c, red curve). When moving the skrymion core towards the Tc-inatom, the non-collinearity leads to the spin-mixing of Fe states. The latter gives rise to additional hybridization channels, which were inactive for V-inatom. That broadens the Tc-minority band occupying the bonding states (see upper arrow in Figure~\ref{Fig3}d, grey curve). This provides a stabilization mechanism of the non-collinear structure at the vicinity of Tc. Among the 4d series, Tc has the largest occupation of the minority bonding state, promoting it to be the strongest pinning center. For instance, the bonding state is less occupied for Mo while for Ru the minority-spin anti-bonding state gets partly populated, lowering in both cases the stability of a skyrmion at their vicinity in comparison to Tc. At the end of the 4d series, both states are occupied and therefore Rh does not interact much with the skyrmion.

Similar arguments can be used for all the investigated impurities (see Supplementary Note 2). In Figure~\ref{Fig2}c-d, we provide for completeness the results obtained for the adatoms. The minimum in the energy-profile is shifted towards Fe (Ru) for the 3d-atoms (4d-atoms). Since adatoms are located further from the Fe-layer than inatoms, the lower direct hybridization leads to a smaller bandwidth of the binding energies (minimum to maximum) compared to the one for inatoms. Interestingly, our results indicate that the Co-adatom is inert, which explains the non-ability of recent STM experiments to use it for skyrmions manipulation\cite{Hanneken2016}. Instead of the Co-adatom, good candidates for atomic control of skyrmions are V, Cr, Mn, Fe, Ru and Rh adatoms. The results are summarized as part of the periodic table of skyrmion-defect interaction, Figure~\ref{Fig2}e, with the colour scale indicating the strength of the binding energies.

\section*{Discussion}

We have systematically studied realistic single magnetic skyrmions in PdFe/Ir(111) surface interacting with 3d and 4d single atomic defect completely from first principles. We have established the energy-profile for the skyrmion-defect interaction as a function of the chemical nature of the defect and related it to the filling of bonding and anti-bonding electronic states inherent to the proximity of the non-collinear magnetic structure. The energy-profile displays a typical signature of a electronic-filling effect and its similarities with the universal trend of cohesion or surface energies of transition metals suggests the universality of the interaction profile. We conjecture that such an effect should characterize the energy-barriers for the nucleation or collapse of magnetic skyrmions, which is fundamentally and technologically important in the field of skyrmionics. The evaluation of such energy-barriers requires, however, to prospect all possible configurational paths, using for example the geodesic nudged elastic band method~\cite{Bessarab2015}, separating two states of different topological nature.

\begin{figure}
	\includegraphics[width=\columnwidth]{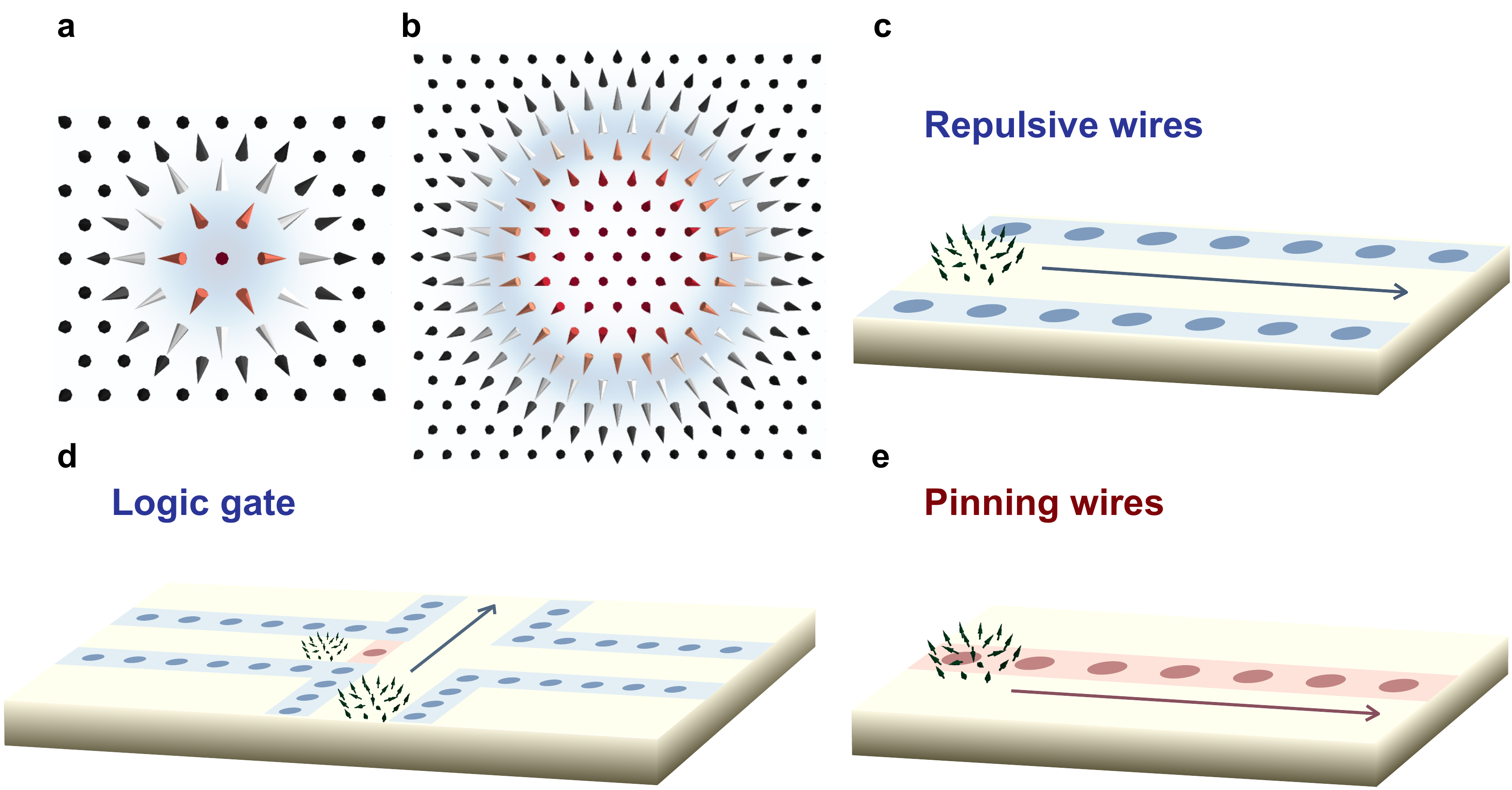} 
	\caption{\textbf{Spintronic devices with defects guiding the skyrmion motion.} \textbf{(a-b)} Non-collinear regions (blue area) defining pinning or repulsion depending on the chemical nature of the impurities. A pinning impurity can lock a skyrmion within the blue region while a repulsive impurity can pin a large skyrmion on the core region, which is a local minimum energy. \textbf{(c-e)} Combination of different energy-landscapes to engineer spintronics devices. \textbf{(c)} Repulsive lanes (blue) can confine the motion of skyrmions in a desired direction. \textbf{(d)} Pinning (red) usually hinders skyrmion mobility and can be used to decelerate them. \textbf{(e)} Pinning lanes (red) can be a medium for skyrmions motion.}
	\label{Fig4}
\end{figure}

Our investigations show that attraction or repulsion of skyrmions and defects is related to the region of high non-collinearity (Figure \ref{Fig4}a-b)  since the gain or loss in energy is related to the spin-mixing hybridization. For large skyrmions the regions of high non-collinearity is shifted away from the skyrmion core, so the pinning or repulsion can occur away from the center. This can lead to interesting local energy minima. The energy-well acquires then a ring-like structure, likewise for the energy-barrier, where the skyrmion core defines a local minimum energy (see blue shadow in Figure \ref{Fig4}b). Thus, Tc-inatom can pin a large skyrmion within the mentioned ring-like region while V-inatom, being initially repulsive, can by accident trap a skyrmion if it is located in the central region. This way, a repulsive impurity can become a pinning center. This gives rise to a plethora of possibilities to shape the energy-profiles of skyrmions, which would certainly ignite interesting and unforeseen internal dynamical effects during the motion. The impact on the recently defined motion-regimes of skyrmions upon application of a current is an open question. In Figure \ref{Fig4}c-e, we propose to design surfaces or interfaces of materials with patterned defects, via atomic manipulation or ion-implantation, to construct highways for skyrmion transport. Single atomic defects can be arranged in lines defining energy-profiles that can conduct skyrmions in well-defined directions. Employing the skyrmion-defect interaction table and its universal nature, one can combine the energy landscape to be produced by different defects to engineer new energy profiles for the skyrmion control and functionality at will. Also, one can predict which impurities to use for the design of skyrmion--pinning or --repulsive centers. The latter ones can be arranged in wires to define energy-local-minimum roads for skyrmion transport whereas even pinning centers, once wisely arranged, can be useful to guide or to decelerate skyrmions for reading processes or for non-trivial multiple-skyrmions motion. We note that other theoretical proposals for skyrmions racetrack were recently discussed (see e.g.~\cite{Fert2013,Stosic2017,Zhang2015c,Mueller2017}).

While the current study focused on the correlation between the electronic structure and the interaction of skyrmions with atomic  defects, we expect that the latter would have a non-negligible effect on transport properties such as the spin-mixing resistance effect~\cite{Crum2015,Hanneken2015} and other magnetic properties pertaining to magnetic skyrmions such as the chiral or topological orbital magnetization~\cite{Dias2016}.

\begin{methods}

The calculations are based on density functional theory considering the local spin density approximation. We use the all-electron full-potential  scalar-relativistic Korringa-Kohn-Rostoker (KKR) Green function method with spin-orbit coupling added self-consistently, which allows to seamlessly embed single magnetic skyrmions and the defects in the magnetic substrate (see Ref.~\cite{Crum2015} and references therein).  Besides the Fe atoms defining the skyrmion and the impurity also all nearest neighbours atoms are included which lead to an embedding cluster of 124 atoms. The non-collinear structures are obtained self-consistently, permitting access to the total energy and  the details of the electronic structure of the whole ensemble comprising the skyrmion, defects and the unperturbed magnetic background. The calculations were performed with an angular momentum cutoff of $l_{\text{max}} = 3$ for the orbital expansion of the Green function. The energy contour contained 42 grid points in the upper complex plane with seven Matsubara poles and a Brilloun zone mesh of $30 \times 30$ k-points for the self-consistent description of pure substrate properties. When embedding the skyrmions and defects, the k-mesh was increased to $200 \times 200$ points.

\noindent \textbf{Code availability} The simulation code used to produce the findings based on the extended Heisenberg model is available from the corresponding authors on request.

\noindent \textbf{Data availability} The data that support the findings of this study are available from the corresponding authors on request.

\end{methods}

\section*{References}
\bibliographystyle{naturemag}
\bibliography{references}

\begin{addendum}

\item We acknowledge helpful discussions with P. H. Dederichs, F. S. M. Guimar\~{a}es, J. Chico and J. Iba\~{n}ez-Azpiroz. We thank P. R\"{u}\ss{}mann for optimizing the full-potential relativistic KKR-Green function code. This work is supported by the European Research Council (ERC) under the European Union's Horizon 2020 research and innovation programme (ERC-consolidator grant 681405 — DYNASORE). I. L. F. acknowledges funding from the Brazilian funding agency CNPq under Project No. 233784/2014-4.  S. B. acknowledges funding from the research and innovation programme under grant agreement number 665095 (FET-Open project MAGicSky)  and the DARPA TEE program through grant MIPR $\left(\#\   \text{HR0011831554}\right)$ from DOI.

\item[Author contributions] I. L. F performed the simulations based on \textit{ab initio} and on the extended Heisenberg model. All authors discussed the results and helped writing the manuscript. 

\item[Competing Interests]  The authors declare that they have no competing financial interests. 

\item[Correspondence] Correspondence and requests for materials should be addressed to I.L.F. (email: i.lima.fer\-nandes@fz-juelich.de) or to S.L. (email: s.lounis@fz-juelich.de).

\end{addendum}

\end{document}